\documentclass[prd,aps,twocolumn,nofootinbib]{revtex4-1}

\usepackage{amssymb}
\usepackage{amsmath}
\usepackage{graphicx}
\usepackage[margin=5pt, font=small,labelfont=bf,justification=raggedright]{caption}

\usepackage{url}
\usepackage[bookmarks, pagebackref=false]{hyperref}
\usepackage{color}
	\definecolor{rossoCP3}{cmyk}{0,.88,.77,.40}
		\definecolor{graa}{rgb}{0.8,0.8,0.8}
		\definecolor{blaa}{rgb}{0.2,0.2,0.6}
		\definecolor{gron}{RGB}{0,150,0}
		\hypersetup{
			colorlinks, 
			bookmarksopen, 
			bookmarksnumbered,
			citecolor=blaa, 		
			linkcolor=black, 
			urlcolor=blaa,			
			}

\newcommand{\ea}[1]{
\begin{align}
#1
\end{align}
}

\newcommand{\nn}{\nonumber \\ }
\newcommand{\MSB}{{\overline{\rm MS}}}
\newcommand{\R}{{\cal R}}

\newcommand{\rd}{{\rm d}}

\reversemarginpar

\begin{document}
{\par \texttt{SLAC-PUB-15299}\par}
{\par \texttt{CP3-Origins-2012-032 \& DIAS-2012-33}\par}
\bigskip{}

\title{A Systematic All-Orders Method to Eliminate Renormalization-Scale and Scheme Ambiguities in PQCD}

\author{Matin Mojaza}
\email{mojaza@cp3-origins.net}
\affiliation{CP3-Origins, Danish Institute for Advanced Studies, University of Southern Denmark, DK-5230 \\
SLAC National Accelerator Laboratory, Stanford University, Stanford, California 94039, USA}

\author{Stanley J. Brodsky}
\email{sjbth@slac.stanford.edu}
\affiliation{SLAC National Accelerator Laboratory, Stanford University, Stanford, California 94039, USA}

\author{Xing-Gang Wu}
\email{wuxg@cqu.edu.cn}
\affiliation{Department of Physics, Chongqing University, Chongqing 401331, P.R. China}

\date{\today}

\begin{abstract}

We introduce a generalization of the conventional renormalization schemes used in dimensional regularization, which illuminates the renormalization scheme and scale ambiguities of pQCD predictions, exposes the general pattern of nonconformal $\{ \beta_i\}$ terms, and reveals a special degeneracy of the terms in the perturbative coefficients. It allows us to systematically determine the argument of the running coupling order by order in pQCD in a form which can be readily automatized. The new method satisfies all of the principles of the renormalization group and eliminates an unnecessary source of systematic error.

\begin{description}
\item[PACS numbers] 12.38.Aw, 12.38.Bx, 11.10.Gh, 11.15.Bt
\end{description}
\end{abstract}

\maketitle

An important goal in high energy physics is to make perturbative QCD (pQCD) predictions as precise as possible, not only to test QCD itself, but also to expose new physics beyond the standard model. In this letter we present a systematic method which  determines the argument of the running coupling order by order in pQCD and which can be readily automatized. The resulting predictions for physical processes are independent of theoretical conventions such as the choice of renormalization scheme and the initial choice of renormalization scale. The resulting scales also  determine the effective number of quark flavors at each order of perturbation theory. The method can be applied to processes with multiple physical scales and is consistent with QED scale setting in the limit $N_c \to 0.$ The new method satisfies all of the principles of the renormalization group \cite{pmcproperties}, and it eliminates an unnecessary source of systematic error.

The starting point for our analysis is to introduce a generalization of the conventional  schemes used in dimensional regularization in which a constant $-\delta$ is subtracted in addition to the standard  subtraction  \mbox{$\ln 4 \pi - \gamma_E$} of the $\MSB$-scheme. 
This amounts to redefining the renormalization scale by an exponential factor; i.e. 
\mbox{$\mu_\delta^2 = \mu_\MSB^2 \exp(\delta)$}. In particular, the MS-scheme is recovered for \mbox{$\delta = \ln 4 \pi -  \gamma_E$}. The $\delta$-subtraction defines an infinite set of renormalization schemes which we call \mbox{$\delta$-$\cal R$enormalization ($\R_\delta$)} schemes; since physical results cannot depend on the choice of scheme, predictions must be independent of $\delta$. Moreover, since all $\cal R_\delta$ schemes are connected by scale-displacements, the $\beta$-function of the strong QCD coupling constant $a = \alpha_s/4\pi$ is the same in any $\R_\delta$-scheme:
\ea{
\label{beta}
 \mu_\delta^2\frac{d a}{d\mu_\delta^2} = \beta(a) = - a(\mu_\delta)^2 \sum_{i=0}^\infty \beta_i a(\mu_\delta)^i \ .
}

The $\R_\delta$-scheme exposes the general pattern of nonconformal $\{\beta_i\}$-terms, and it reveals a special degeneracy of the terms in the perturbative coefficients which allows us to resum the perturbative series. The resummed series matches the conformal series, which is itself free of any scheme and scale ambiguities
as well as being free of a divergent ÒrenormalonÓ series. It is the final expression one should use for physical predictions. It also makes it possible to set up an algorithm for automatically computing the conformal series and setting the effective scales for the coupling constant at each perturbative order.

Consider an observable in pQCD in some scheme which we put as the reference scheme $\R_0$ (e.g. the $\MSB$-scheme, which is the conventional scheme used) with the following expansion:
\ea{
\rho_0 (Q^2)= \sum_{i=0}^\infty r_i (Q^2/\mu_0^2) a(\mu_0) ^i \ ,
}
{where $\mu_0$ stands for the initial renormalization scale and $Q$ is the kinematic scale of the process}. The more general expansion with higher Born-level power in $a$ can be readily derived \cite{BMW} and will not change our conclusions and results. 
The full pQCD series is formally 
 independent of the choice of the initial renormalization scale $\mu_0$, 
 if it were possible to sum the entire series.
 However, this goal is not feasible in practice, especially because of the $n! \beta^n \alpha_s^n$ renormalon growth of the nonconformal series. When a perturbative expansion is truncated at any finite order, it generally becomes renormalization-scale and scheme dependent; i.e.,  dependent on theoretical conventions.
This can be exposed by using the $\R_\delta$-scheme.
Since results in any $\R_\delta$ are related by scale displacements, we can derive a general expression for $\rho$ by using the displacement relation between couplings in any $\R_\delta$-scheme:
\ea{
\label{adelta}
a(\mu_0) = a(\mu_\delta) + \sum_{n=1}^\infty \frac{1}{n!} { \frac{{\rm d}^n a(\mu)}{({\rm d} \ln \mu^2)^n}\left|_{\mu=\mu_\delta}\right. (-\delta)^n} \ ,
}
{where we used $\ln\mu^2_0/\mu^2_\delta=-\delta$}. Then $\rho$ in $\R_\delta$ to order $a^4$ reads:
\ea{
\label{rhodeltas}
\rho_\delta(Q^2) = &r_0 + r_1 a_1(\mu_1) + (r_2 + \beta_0 r_1 \delta_1) a_2(\mu_2)^2 \nn
& + [r_3 +\beta _1 r_1 \delta_1+  2 \beta _0 r_2\delta_2+ \beta _0^2 r_1 \delta_1^2 ] a_3(\mu_3)^3 \nn
& +[r_4 +\beta _2 r_1\delta_1 +2 \beta _1 r_2\delta_2  +3 \beta _0 r_3\delta_3 +3 \beta _0^2 r_2\delta_2^2
\nn
&+\beta _0^3 r_1 \delta_1 ^3 +\frac{5}{2} \beta _1 \beta _0 r_1\delta_1^2 ] a_4(\mu_4)^4 +{\cal O}(a^5) \ .
}
where $\mu_i^2 = Q^2 e^{\delta_i}$, the initial scale is for simplicity set to $\mu_0^2 = Q^2$ and we defined $r_i(1) = r_i$. An artificial index was introduced on each $a$ and $\delta$ to keep track of which coupling each $\delta$ term is associated with. The initial scale choice is arbitrary and is not the final argument of the running coupling;
the final  scales will be independent of the initial renormalization scale.

In a  conformal (or scale-invariant) theory, where $\{\beta_i\}=\{0\}$, the $\delta$ dependence vanishes in Eq.\eqref{rhodeltas}. 
Therefore by absorbing all $\{\beta_i\}$ dependence into the running coupling at each order, we obtain a final result independent of the initial choice of scale and scheme. 
The use of $\R_\delta$ allows us to put this on rigorous grounds.
From the explicit expression in Eq.\eqref{rhodeltas} it is easy to confirm that
\begin{align}
\frac{\partial \rho_\delta}{\partial \delta} = -\beta(a) \frac{\partial \rho_\delta}{\partial a} \ .
\end{align}
The scheme-invariance of the physical prediction requires that $\partial \rho_\delta /\partial \delta = 0$. 
Therefore the scales in the running coupling must be shifted and set such that the conformal terms associated with the $\beta$-function are removed;  the remaining conformal terms are by definition renormalization scheme independent.
The numerical value for the prediction at finite order is then scheme independent as required by the renormalization group.
The scheme-invariance criterion is a theoretical requirement of the renormalization group; it must be satisfied at any truncated order of the pertubative series, and is different from the formal statement that the all-orders expression for a physical observable is renormalization scale and scheme invariant; i.e. ${\rm d} \rho /{\rm d} \mu_0 = 0$.
The final series obtained corresponds to the theory for which
\mbox{$\beta(a) = 0$}; i.e. the conformal series.
This demonstrates to any order the concept of the \emph{principal of maximum conformality} \cite{pmc1,pmcprl} (PMC), which states that all non-conformal terms in the perturbative series must be resummed into the running coupling.

The expression in Eq.\eqref{rhodeltas} exposes the pattern of $\{\beta_i\}$-terms in the coefficients at each order. Such a pattern was recently considered in Ref.~\cite{Mikhailov}. 
The $\R_\delta$-scheme reveals its origin and 
its generality for any pQCD prediction. 
It is possible to infer more from Eq.\eqref{rhodeltas}; since
 there is nothing special about a particular value of $\delta$,
we conclude that some of the coefficients of the $\{\beta_i\}$-terms are degenerate; 
e.g. the coefficient of $\beta_0 a(Q)^2$ and $\beta_1 a(Q)^3$ can be set equal.
Thus for any scheme, the expression for $\rho$ can be put to the form:
\ea{
\rho(Q^2) = &r_{0,0} + r_{1,0} a(Q) + [r_{2,0} + \beta_0 r_{2,1} ] a(Q)^2   \nn
&+[r_{3,0} + \beta_1 r_{2,1} + 2 \beta_0 r_{3,1} + \beta _0^2 r_{3,2} ]a(Q)^3  \nn
&+[r_{4,0} + \beta_2 r_{2,1}  + 2\beta_1 r_{3,1} +  \frac{5}{2} \beta_1 \beta_0 r_{3,2} +3\beta_0 r_{4,1}\nn
& \quad+ 3 \beta_0^2 r_{4,2} + \beta_0^3 r_{4,3} ] a(Q)^4  +{ \cal O }(a^5)  \label{betapattern}
}
where the $r_{i,0}$ are the conformal parts of the perturbative coefficients; i.e. 
$r_{i} = r_{i,0} + {\cal O}(\{\beta_i\})$. 
The $\R_\delta$-scheme not only illuminates the $\{\beta_i\}$-pattern, but it also exposes a \emph{special degeneracy} of coefficients at different orders.
The degenerate coefficients can themselves be functions of $\{\beta_i\}$, hence
Eq.\eqref{betapattern} is not to be understood as an expansion in $\{\beta_i\}$, but
a pattern in $\{\beta_i\}$ with degenerate coefficients that must be matched.
We have checked that this degeneracy holds for several known results.

The expansion in Eq.\eqref{rhodeltas} reveals how the $\{\beta_i\}$-terms must be absorbed into the running coupling. The different $\delta_k$'s keep track of the 
power of the $1/\epsilon$ divergence of the associated diagram at each loop order in the following way; the $\delta_k^p a^n$-term indicates the term associated to 
a diagram with $1/\epsilon^{n-k}$ divergence for any $p$.
Grouping the different $\delta_k$-terms, one recovers
in the $N_c \to 0$ Abelian limit \cite{Brodsky:1997jk} the dressed skeleton expansion.
Resumming the series according to this expansion thus
correctly reproduces the QED limit of the observable
and matches the conformal series with running coupling constants
evaluated at effective scales at each order.

Using this information from the $\delta_k$-expansion, 
it can be shown that the order 
$a(Q)^k$ coupling must be resummed
into the effective coupling $a(Q_k)^k$, given by:
\begin{widetext}
\ea{
r_{1,0} a(Q_1) &= r_{1,0} a(Q) - \beta(a) r_{2,1} + \frac{1}{2} \beta(a) \frac{\partial \beta}{\partial a} r_{3,2} + \cdots + \frac{(-1)^n}{n!} \frac{\rd^{n-1}\beta}{(\rd \ln\mu^2)^{n-1}} r_{n+1,n} \ , \label{1order} \\[-3mm]
&  \hspace{2mm} \vdots \nonumber \\[-3mm]
r_{k,0} a(Q_k)^{k} &= r_{k,0} a(Q)^k + r_{k,0}\  k \ a(Q)^{k-1} \beta(a) \left \{
R_{k,1} +\Delta_k^{(1)}(a) R_{k,2} + \cdots + \Delta_k^{(n-1)}(a) R_{k,n} \right \} \ ,
\label{korder}}
which defines the PMC scales $Q_k$ and where we introduced
\ea{
R_{k,j} = (-1)^{j}\frac{r_{k+j, j}}{r_{k,0}} \ , \quad \Delta_k^{(1)}(a) = \frac{1}{2} \left [ \frac{\partial \beta}{\partial a} + (k-1) \frac{\beta}{a}\right]  \ , \quad \cdots
}
Eq.\eqref{korder} is systematically derived by replacing the $\ln^j Q_1^2/Q^2$ by $R_{k,j}$ in the logarithmic expansion of $a(Q_k)^k$ up to the highest known $R_{k,n}$-coefficient in pQCD.
The resummation can be performed iteratively using the renormalization group equation for $a$ and leads to the effective scales for an NNNLO prediction\footnote{Detailed derivations will be given elsewhere \cite{BMW}}:
\ea{
\label{exactscales}
\ln \frac{Q_{k}^2}{Q^2}  &= \frac{R_{k,1} + \Delta_k^{(1)}(a) R_{k,2}+\Delta_k^{(2)}(a) R_{k,3}}{1+ \Delta_k^{(1)}(a) R_{k,1} + \left({\Delta_k^{(1)}(a)}\right)^2 (R_{k,2} -R_{k,1}^2)  + \Delta_k^{(2)}(a)R_{k,1}^2 } \ .
}
The final pQCD prediction for $\rho$ after setting the PMC scales $Q_i$ then reads
\ea{
\rho(Q^2) = &r_{0,0} + r_{1,0} a(Q_1) + r_{2,0} a(Q_2)^2 +r_{3,0} a(Q_3)^3 + r_{4,0}a(Q_4)^4+{ \cal O }(a^5) \ ,
}
\end{widetext}
Note that $Q_4$ remains unknown.
This last ambiguity resides only in the highest order coupling constant, and is negligible in practice.

It is easy to see to leading logarithmic order (LLO) that the effective scales are independent of the initial renormalization scale $\mu_0$.
This follows since taking $\mu_0 \neq Q$ we must replace $R_{k,1} \to R_{k,1} + \ln Q^2/\mu_0^2$ and thus the leading order effective scales read
$\ln Q_{k, \rm LO}^2/\mu_0^2 = R_{k,1} + \ln Q^2/\mu_0^2$, where $\mu_0$ cancels and Eq.~\eqref{exactscales} at LLO is recovered. This generalizes to any order. 
In practice, however, since the $\beta$-function is not known to all orders, 
 a higher order residual renormalization-scale dependence will enter through the running coupling constant. This residual renormalization-scale dependence is strongly suppressed in the perturbative regime of the coupling \cite{pmctop1}.

The effective scales contain all the information of the non-conformal parts of the initial pQCD expression for $\rho$ in Eq.\eqref{betapattern}, 
which is exactly the purpose of the running coupling constant.
The quotient form of Eq.~\eqref{exactscales} sums up an infinite
set of terms related to the known $r_{j,k \neq 0}$ which appear at every higher order due to the special degeneracy. It is, however, not the full solution since this requires the knowledge of the $r_{j,k \neq 0}$-terms to all orders. The method systematically sums up all known non-conformal terms, in principle to all-orders, but is in practice truncated due to the limited knowledge of the $\beta$-function.

In earlier PMC scale setting \cite{pmcprl}, 
and its predecessor, the Brodsky-Lepage-Mackenzie (BLM) method \cite{blm},
the PMC/BLM scales have been set by using a perturbative expansion in $a$ and only approximate conformal series have been obtained. Here, we have been able to 
obtain the conformal series exactly due to the revelation of the $\{\beta_i\}$-pattern by $\R_\delta$; the effective scales have naturally become functions of the coupling constant through the $\beta$-function, in principle, to all orders.

In many cases the coefficients in a pQCD expression for an observable are computed numerically, and
the $\{\beta_i\}$ dependence is not known explicitly. It is, however, easy to extract the dependence on the number of quark flavors $N_f$, since $N_f$ enters analytically in any loop diagram computation.
To use the systematic method presented in this letter one puts the pQCD expression into the form of Eq.\eqref{betapattern}. Due to the special degeneracy in the coefficient of the $\{\beta_i\}$-terms, the $N_f$ series 
can be matched to the $r_{j,k}$ coefficients in a unique way%
\footnote{
In principle, one must treat the $N_f$  terms unrelated to renormalization of the gauge coupling as part of the conformal coefficient.
}.
This allows one to automate the
scale setting process algorithmically.

The general $N_f$-series of the $n$-th order coefficient in pQCD reads:
\ea{
r_{n} = c_{n,0} + c_{n,1} N_f + \cdots + c_{n,n-1} N_f^{n-1} \ .
}
By inspection of Eq.\eqref{betapattern} it is seen that there are exactly as many unknown coefficients in the $\{\beta_i\}$-expansion at the order $a^n$ as the $N_f$ coefficients, $c_{n,j}$. This is realized due to the special degeneracy found in \eqref{betapattern}.
The $r_{i,j}$ coefficients in Eq.\eqref{betapattern} can thus be expressed in terms of the $c_{n,j}$ coefficients. This means that the $N_f$ terms can unambiguously be associated to $\{\beta_i\}$-terms and demonstrates PMC as the underlying principle of BLM scale setting. The relations between $c_{n,j}$ and $r_{i,j}$ are easy to derive and they transform the BLM scales into the correct PMC scales \cite{BMW}.

The automation process can be outlined as follows:
\begin{enumerate}
\item Choose any $\delta$-$\cal R$enormalization scheme and scale.
\item Compute the physical observable in pQCD and extract the $N_{\rm f}$ coefficients, $c_{k,j}$. 
\item Find the $\beta_i$ coefficients, $r_{k,j}$ from the $c_{k,j}$ coefficients and compute the PMC scales, $Q_k$.
\item The final pQCD expression for the observable reads $\rho_{\rm final} (Q) = \sum_{k=0} r_{k,0} a(Q_k)^{k}$.
\end{enumerate}

As a final remark, we note that
the PMC can be used to set separate scales for different skeleton diagrams; this is particularly important for multi-scale processes. In general the $\{\beta_i\}$-coefficients multiply terms involving logarithms in each of the invariants \cite{pmc1}.
For instance, in the case of $q \bar q \to Q \bar Q$ near the heavy quark threshold in pQCD, the PMC assigns different scales to the annihilation process and the rescattering corrections involving the heavy quarks' relative velocity \cite{Brodsky:1995ds}.
It also can be used to set the scale for the ``lensing" gluon-exchange corrections that appear in the Sivers, Collins, and Boer-Mulders effects.
Moreover, for the cases when the process involves several energy regions; e.g. hard, soft, etc., one may adopt methods such as the non-relativistic QCD effective theory (NRQCD) \cite{NRQCD} and the soft-collinear effective theory (SCET) \cite{SCET} to set the PMC scales; i.e., one first sets the PMC scales for the higher energy region, then integrate it out to form a lower energy effective theory and sets the PMC scales for this softer energy region, etc. In this way one obtains different effective PMC scales for each energy region, at which all the PMC properties also apply. 

\noindent{\bf Example: $e^+e^- \to$ {hadrons}.} The ratio for electron-positron annihilation into hadrons, $R^{e^+e^- \to \textbf{h}}$,
 was recently computed to order $a^4$ \cite{Baikov:2012zn}
and can be shown to exactly match the generic form of Eq.\eqref{betapattern}. It can be derived by analytically continuing the Adler function, $D$, into the time-like region, with $D$ given by:
\ea{
D(Q^2) &=  \gamma(a) - \beta(a) \frac{d}{da} \Pi(Q^2,a) \ ,
}
where $\gamma$ is the anomalous dimension of the vector field, $\Pi$ is the vacuum polarization function and they are given by the perturbative expansions: \mbox{$\gamma (a) = \sum_{n=0}^\infty \gamma_n a^n$} and \mbox{$\Pi(a)  = \sum_{n=0}^\infty \Pi_{n} a^n$}. It is easy to show that to order $a^4$ the perturbative expression for $R^{e^+e^- \to \textbf{h}}$ in terms of $\gamma_n$ and $\Pi_n$ reads:
\ea{
\label{Ree}
R^{e^+e^- \to \textbf{h}}&(Q) = \gamma_0 + \gamma_1 a(Q) + [\gamma_2 + \beta_0 \Pi_1] a(Q)^2 \\
& + [\gamma_3 + \beta_1 \Pi_1 + 2 \beta_0 \Pi_2- \beta_0^2 \frac{\pi^2\gamma_1}{3}  ] a(Q)^3 \nn
& + [\gamma_4 + \beta_2 \Pi_1 + 2 \beta_1 \Pi_2 + 3 \beta_0 \Pi_3 \nn
&- \frac{5}{2} \beta_0\beta_1 \frac{\pi^2\gamma_1}{3} - 3 \beta_0^2 \frac{\pi^2 \gamma_2}{3} -  \beta_0^3 \pi^2 \Pi_1 ] a(Q)^4  \ . \nonumber
}
This expression has exactly the form of Eq.\eqref{betapattern} with the identification:
\mbox{$r_{i,0} = \gamma_i$, } \mbox{$r_{i,1} = \Pi_{i-1} $,} \mbox{$r_{i,2} = - \frac{\pi^2}{3} \gamma_{i-2}$} and \mbox{$r_{i,3} = -\pi^2 \Pi_{i-3}$}. 
The $\gamma_i$ contain $N_f$-terms, but since they are independent of $\delta$ to any order, they are kept fixed in the scale-setting procedure.
Note that we have knowledge of even higher order $r_{i,j}$ coefficients, and this allows us to set the effective scales $Q_1$, $Q_2$ and $Q_3$ to the NNNLO, given by Eq.\eqref{exactscales}.
It is worth noting that the Adler function $D$ itself has a much simpler $\{\beta_i\}$-structure.
By convention the argument of $a$ is space-like; thus the $\pi^2$-terms appearing in $R^{e^+e^- \to \textbf{h}}$ could be avoided by
using a coupling constant with time-like argument, leading to a more convergent series
\cite{pennington}.

The last unknown scale in Eq.~\eqref{Ree} can be estimated.
It turns out that $Q_4 \sim Q $ which is the value we have used \cite{BMW}. The expressions for the coefficients $\gamma_i$ and $\Pi_i$ can be found in Ref.~\cite{Baikov:2012zn}, and the four-loops $\beta$-function is given in Ref.~\cite{vanRitbergen:1997va}.
The final result in numerical form in terms of $\alpha = \alpha_s/\pi$ for QCD with five active flavors reads:
\ea{
\label{Rnum}
\frac{3}{11}  R^{e^+e^- \to \textbf{h}}(Q) =& 1 + \alpha(Q_1) + 1.84 \alpha(Q_2)^2
\nn & - 1.00 \alpha(Q_3)^3 -
 11.31 \alpha(Q_4)^4 \ .
}
This is a more convergent result compared to previous estimates, and it is free of any scheme and scale ambiguities (up to strongly suppressed residual ones).

\begin{figure}[!b]
\vspace{-3mm}
\includegraphics[width=1\columnwidth]{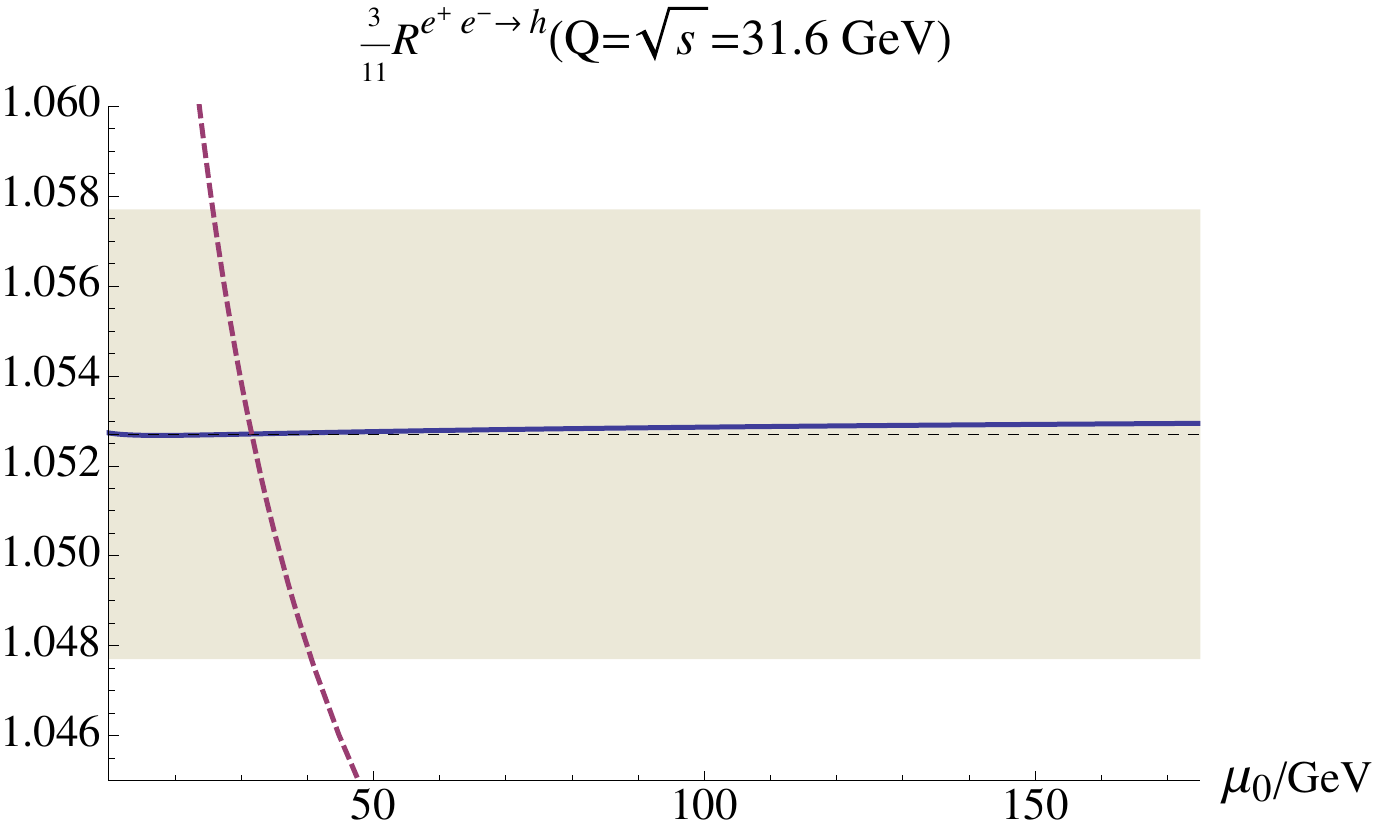}
\caption{The final PMC result for $R^{e^+e^- \to \textbf{h}}$ as a function of the initial renormalization scale $\mu_0$ (solid blue line), demonstrating the initial scale-invariance of the final prediction, up to strongly suppressed residual dependence. The shaded region is the experimental bounds with the central value given by the thin dashed line. 
For comparison we also show the pQCD prediction before PMC scale setting (dashed red line) fixed to the experimental value for $\mu_0 = Q$. This result is very sensitive to $\mu_0$, and thus it severely violates the renormalization group properties.
}
\label{Rsmu}
\end{figure}

To find numerical values for the effective scales, the asymptotic scale, $\Lambda$, of the running coupling must first be determined by matching Eq.\eqref{Rnum} with experimental results
\cite{Marshall:1988ri}:
$\frac{3}{11} R_{\rm exp}^{e^+e^- \to \textbf{h}}(\sqrt{s} = 31.6 \text{ GeV}) = 1.0527 \pm 0.0050 $ .
Using a logarithmic expansion solution of the renormalization group equation for $a$ we find: \mbox{$\Lambda_{\MSB} = 419^{+222}_{-168} \text{ MeV}$}.
We have used the $\MSB$ definition for the asymptotic scale, and the asymptotic scale of $\R_\delta$ can be taken to be the same for any $\delta$. 
The effective scales are found to be: $Q_1 = 1.3 ~Q~, Q_2 = 1.2 ~Q$, $Q_3 \approx 5.3 ~Q$. The values are independent of the initial renormalization scale up to some residual dependence coming from the truncated $\beta$-function, which is less than the quoted accuracy on the numbers. This is illustrated in Fig.~\ref{Rsmu}. For $Q_3$ we have taken the LO value, which is sufficient to get the conformal series at four loops. Its higher order value has artificial strong residual renormalization-scale dependence due to the large numerical value of $\Pi_3$ in QCD with five active flavors.
These final scales determine the effective number of quarks flavors at each order of perturbation theory \cite{Brodsky:1998mf}. 

For completeness, we use our final result to predict the strong coupling constant at the Z-boson mass-scale in five flavor massless QCD:
\begin{align}
\alpha_s(M_Z) = 0.132^{+0.010}_{-0.011}  \ .
\end{align}
The error on this result is a reflection of the experimental uncertainty on
$R_{\rm exp}^{e^+e^- \to \textbf{h}}$, which cannot be eliminated.
This value is somewhat larger than the present world average $\alpha_s(M_Z) =0.1184 \pm 0.0007 $, which is a global fit of all types of experiments. However, it is consistent with the values obtained from $e^+ e^-$ colliders, i.e. $\alpha_s(M_Z)=0.13\pm 0.005\pm0.03$ by the CLEO Collaboration \cite{CLEO} and $\alpha_s(M_Z)=0.1224\pm 0.0039$ from the jet shape analysis \cite{jet}. Moreover, in computing $\alpha_s(M_Z)$ we have assumed massless quarks. The estimate will decrease when taking threshold effects properly into account as shown in \cite{threshold}.

We can apply our result to Abelian QED, where $R^{e^+e^- \to \textbf{h}}$ can be seen as the imaginary part of the QED four loop 1PI vacuum polarization diagram by the optical theorem, and find in this case nearly complete renormalization scale independence of all three scales to the NNNLO due to the small value of the coupling constant. Numerically, we get for three (lepton) flavors:
\ea{
\label{RnumQED}
\frac{1}{3}  R^{e^+e^- \to \ell}_{\rm QED}(Q) =& 1 + 0.24\alpha_e(Q_1) -0.08 \alpha_e(Q_2)^2
\nn & - 0.13 \alpha_e(Q_3)^3 + 0.05 \alpha_e(Q_4)^4 \ ,
}
where $\alpha_e = e^2/4\pi$ and 
$\{\frac{Q_1}{Q},\frac{Q_2}{Q},\frac{Q_3}{Q}\} = \{1.1, 0.6, 0.5\}$.

In this letter we have shown that a generalization of the conventional $\MSB$-scheme is illuminating. It enables one to determine the general (and degenerate) pattern of nonconformal $\{\beta_i\}$-terms and to systematically determine the argument of the running coupling order by order in pQCD, in a way which is readily automatized. The resummed series matches the conformal series, in which no factorially divergent $n! \beta^n \alpha_s^n$ ``renormalon" series appear 
and which is free of any scheme and scale ambiguities. This is the final expression one should use for physical predictions. The method can be applied to processes with multiple physical scales and is consistent with QED scale setting in the limit $N_C \to 0.$ The new method satisfies all of the principles of the renormalization group, including  the principle of maximum  conformality, and it eliminates an unnecessary  source of systematic error.

{\bf Acknowledgements:} 
We thank Joseph Day, Leonardo Di Giustino and Stefan H\" oche for useful discussions. We are grateful to Konstantin Chetyrkin and Andrei L. Kataev for useful comments.
MM thanks SLAC theory group for kind hospitality. This work was supported in part by the Department of Energy contract DE-AC02-76SF00515, the Natural Science Foundation of China under Grant NO.11275280 and the Danish National Research Foundation, grant no. DNRF90.

\end{document}